# Electric-field-induced topological changes in multilamellar and in confined lipid membranes

Di Jin*, Yu Zhang and Jacob Klein*

Dept. of Molecular Chemistry and Materials Science, Weizmann Institute of Science, Rehovot, Israel

**Abstract**

It is well known that lipid membranes respond to a threshold transmembrane electric field through a reversible mechanism called electroporation, where hydrophilic water pores form across the membrane, an effect widely used in biological systems. The effect of such fields on interfacially-confined (stacked or supported) lipid membranes, on the other hand, which may strongly modulate interfacial properties, has not to our knowledge been previously studied. Motivated by recent surface forces experiments showing a striking effect of electric fields on lubrication by confined lipid bilayers, we carried out all-atom molecular dynamics simulations of such membranes under transverse electric fields. We find that in addition to electroporation, a new feature emerges of locally merged bilayers which act to bridge the confining interfaces. These features shed light on the remodelling of confined lipid membrane stacks by electric fields, and provides insight into how such fields may modulate frictional and more generally surface interactions in the presence of lipid-based boundary layers.

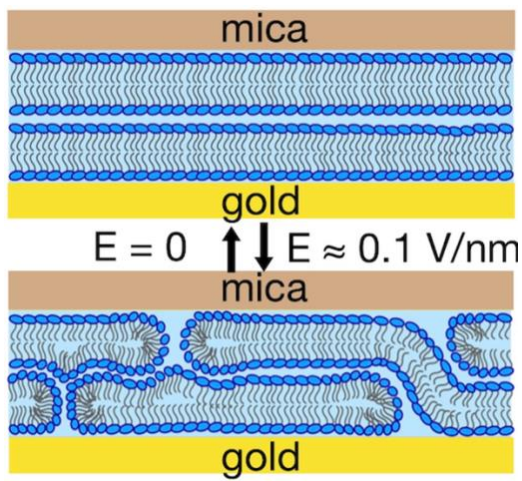

## 1. Introduction

In an aqueous environment, membranes of phosphatidylcholine (PC) lipids (the most common lipid type in living systems) maintain their structural integrity by hydrophobic interactions, in which the inner hydrophobic tail region is shielded by a continuous surface composed of hydrophilic headgroups densely packed with an area per lipid at approximately 60-70 Å$^2$. Even when confined between solid surfaces under high mechanical pressure up to O(100) atm and shear, lipid membrane stacks generally remain separated due to hydration repulsion between the highly-hydrated phosphocholine headgroup layers [1-4]. The fluidity of the associated hydration layers results in extremely low inter-membrane friction (with friction coefficients μ = [(force to slide)/load] down to ca. 10$^{-4}$) as demonstrated by surface force balance (SFB) experiments, and has strong relevance to biological lubrication and related pathologies such as osteoarthritis [5-8]. A recent SFB study [9] reported that on applying a transmembrane electric field of ca. 0.1 V/nm across two similarly-prepared lipid bilayers supported on solid substrates, the friction force between the sliding substrates increases, strikingly, by more than two orders of magnitude, and that the process is reversible. That is, μ changes reversibly from O(5x10$^{-4}$) to O(0.1) on switching on the field, and back again on switching it off. This result demonstrates directly that the electric-field induces reversible changes in the two contacting bilayers that strongly regulate the energy dissipation as they slide past each other, providing massive electro-modulation of the friction, with a dynamic range that is orders of magnitude larger than previously reported [9].

It is well known that PC lipid membranes electroporate above a threshold transmembrane potential of ca. 0.1 V, and do so in a reversible manner [10-16]. A widely accepted explanation of the mechanism is that it is initiated by the interfacial water dipoles aligned at the lipid headgroups, which stochastically penetrate the acyl tail region of the membrane. Lipid headgroups then reorient towards the interior water core until a stable hydrophilic pore forms. Reversibility is demonstrated by the resealing of stabilized hydrophilic pores when the electric field is lifted [10, 14, 16-22]. Despite all the known details of electroporation, the specific structural transformation at bilayer-bilayer interfaces due to electroporation has not been discussed since most of the MD simulations were based on a single bilayer. Electroporation of double bilayers has been previously studied, but mostly motivated by conceptually proving the charge-imbalance method, a method conventionally applied to simulate the effect of electric field pulses of low magnitudes and long durations (> ms) [12, 20]. Most of these simulations are performed with excess hydration of more than 30 water/lipid [14, 20, 22-31], which do not capture the scenario of bilayer stacks or of compressed bilayers as in the SFB experiments. In addition, lipid membrane systems under E-fields either supported or confined between solid surfaces, as in lubricating lipid boundary layers which have also been attributed as underlying biological lubrication mechanisms [5-8], have not to our knowledge been previously studied by MD simulations. This also limits our understanding of supported lipid membranes, a common experimental scheme [8, 32-34]. The solid confining surfaces in our study, namely gold and mica emulating the SFB surfaces, are both adhesive to lipid membranes and thus suppress their out-of-plane fluctuations, a crucial consideration for a realistic simulation.

To understand the molecular-level mechanisms responsible for the SFB results, therefore, we develop here a molecular dynamics (MD) model which is designed to match the experimental conditions as faithfully as possible. We observe structural changes that directly relate to two well-established lipid membrane remodelling mechanisms previously observed under different hydration levels— electroporation, and surprisingly, stalk-like membrane merging which in simulations to date only occurs in presence of fusion proteins or at extreme dehydration [35-41]. These mechanisms, previously only studied for unsupported bilayers, are now considered in the context of modification of *interfacial* properties by confined lipid membranes under electric fields, an area of broad potential interest whenever lipids are used to modify interactions between surfaces. In this work we treat the structural/topological changes of PC bilayers in stacks, and under confinement by solid surfaces at 10 atm, similar to experiment conditions. In a separate study [42], we examine the implications of our findings by correlating friction force measurements in the molecular dynamic simulations with the surface force balance experiments.

2. **Results and discussion**

As this is, to our knowledge, a first simulation of supported or confined lipid bilayers under an electric field, we note that choosing a feasible protocol while minimizing artefacts is challenging when solid boundaries are involved. First, the equilibration time is prolonged as molecules are restricted to rearranging across the periodic boundaries only in 2D. Second, the area of the membrane box is fixed at the area of the solid slabs, which hinders the lateral expansion of a porating membrane, potentially imposing artefacts. Lastly, relating the magnitude of the electric field in the MD simulation with that of the experiment requires consideration of the increased membrane dielectric constant during electroporation, as previously discussed by Böckmann's group [10, 17]. In this study, we overcome these issues by developing a protocol involving both the direct electric field method and the charge imbalance method [10, 14, 16-23]. We first equilibrate the electroporated membrane structure in the absence of solid boundaries under direct electric fields with 3D periodic boundary condition, then equilibrate again with fitted solid slabs and under a transmembrane potential imposed by charge-imbalance. The electric fields between the two methods are carefully matched by addressing the average dielectric constant of the lipid-water mix, based on the geometry of the porated membrane.

The range of hydration levels corresponding to the conditions in the SFB experiments are carefully determined. In these, a hard wall distance of 9-10 nm is arrived at an increasing compression in the range of O(1 atm) to O(10 atm), and lubricity is maintained by the water molecules tenaciously bound to the zwitterionic PC headgroups [2, 3, 43]. In this scenario, compression removes free water molecules from the bilayer interface, and the minimal hydration level is assumed. With MD simulation analysis, Murzyn et al. (2001) and Foglia et al. (2010) calculated the sum of the water molecules H-bonded to headgroup oxygens and clathrated around the phosphocholine group to be 11.6-12 water/lipid [44, 45]. It is consistent with 9.4±0.5 water/lipid measured by NMR and neutron diffraction with supported POPC membrane stacks exposed to 93% humidity [46]. Early NMR quadrupole splitting measurements on heavy water also suggested 11-12 water per lipid in the hydration shell [47, 48]. Based on these data, we assumed a hydration level of $\underline{n}_w$ = 12 water/lipid for perfect membranes under compression. In addition, in the SFB experiments, lipid membranes were first deposited on the solid substrates by incubation and then rinsed with pure water to remove loosely attached surface liposomes. The treatment of rinsing introduces submicron water defects, which are readily observed with AFM scans, in which case higher local hydration levels are expected (Figure 1). By analyzing the AFM scans with image segmentation, we estimated an areal ratio of 13% for the water defects, which translates to a mean value $n_w$ = 20 water/lipid (Figure S 1, Figure S 2).

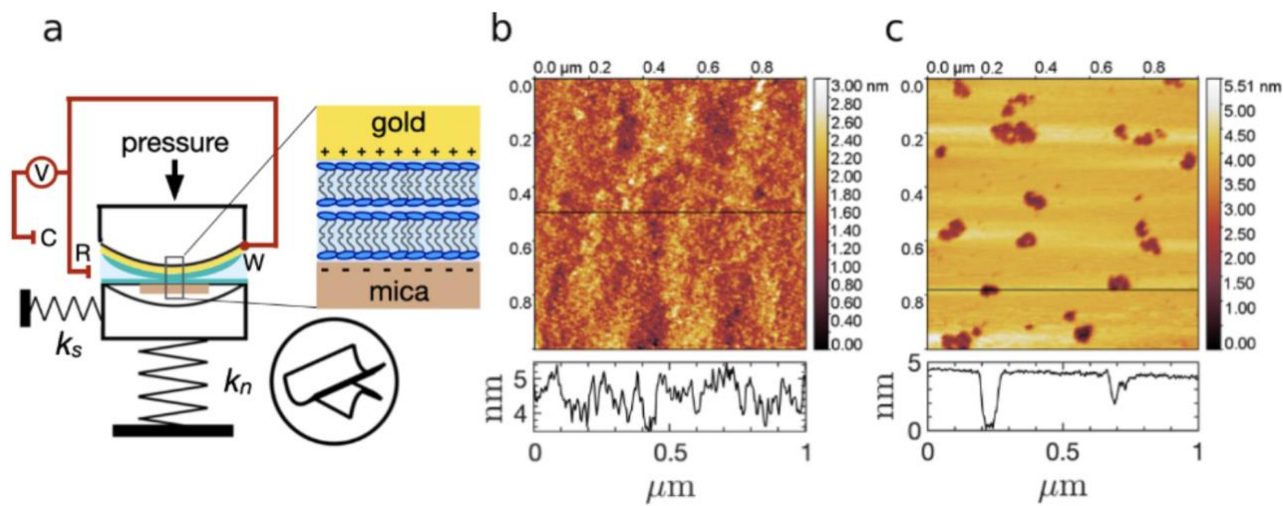

Figure 1. a) Schematics of the surface force balance experiment. Two optical lenses are arranged in a cross-cylinder configuration to ensure a flat contact area. Pressure loaded to the upper lens is measured by the deflection of the normal spring ($K_n$) and the shear force is measured with the shear spring ($K_s$) upon initiation of lateral relative motion with the upper lens. The cleaved mica surface is atomically smooth, while the gold surface, prepared by template-stripping, is of a 0.1-0.2 nm root mean square roughness and functions as the working electrode. (V: potentiostat, C: counter electrode, R: reference electrode, W: working electrode). For details see elsewhere [49-51]. Blue: POPC headgroups. Grey: POPC carbon tails. Light blue: pure water bath. b-c) AFM scans of POPC bilayer on gold and mica surfaces respectively, after incubated in liposome dispersion and rinsed with pure water.

Hydration levels turned out to be an important parameter controlling the inter-bilayer interface, which led to the discovery of two lipid remodelling modes. At minimum hydration level, electroporation introduces topological interbilayer locking due to dehydration and enhanced membrane undulation. At the higher hydration level, sufficient lipid rearrangement enables the formation of hydrophobic contacts, which lead to the formation of vertical lipid bridges connecting the two planar membranes. The later effect leads to interesting comparison with the structurally similar stalk formation—a metastable intermediate state of membrane fusion [35, 38, 52]. Merging membrane stacks requires overcoming hydration repulsion [35, 37, 53]. For stalk formation, this is achieved either through fusion molecules such as SNARE, which locally pinch the membranes together or by exposing the bilayer stack to controlled relative humidity and therefore controlled dehydration [35-39, 41, 52, 54]. Existing MD simulation studies demonstrated that for POPC bilayers, stalk formation occurs spontaneously at 5 water/lipid and at elevated temperature of 380 K or with induced hydrophobic interaction [36]. In a striking difference, our newly discovered mechanism is reversibly induced at room temperature, in an aqueous environment, by a macroscopic and nonintrusive external electric field, applied to the membrane stack with the headgroup hydration shell mostly preserved even under a pressure of O(10) atm. In addition, the membranes remain stably separated with localized merging points without going through the irreversible process of hemifusion as led by the stalk formation.

## 2.1 Transmembrane electric field between charged surfaces across porated bilayers

The mean effective macroscopic transmembrane electric field $E(\epsilon)$ across the bilayers between oppositely charged surfaces is associated with a porated structure and a mean dielectric constant $\epsilon$ of the porated bilayers which differs from that of either the lipid bilayer alone (for which $\epsilon_{\text{bilayer}} \approx 3$) or water ($\epsilon_w \approx 80$) [55]. In this section we self-consistently estimate this electric field, with reference to the SFB experiments. Stable electroporated membranes in the simulation are created with the conventional method: pore formation is induced with a higher electric field of 0.5 V/nm, then the electric field is reduced to a lower value $E$ and equilibrated [10, 12, 14, 21, 56, 57]. Figure 2a shows the equilibrium pore radius of membrane-only systems (i.e. no supporting or confining solid surface) arising from the simulations at different hydration levels as a function of different electric fields (fig. 2b).

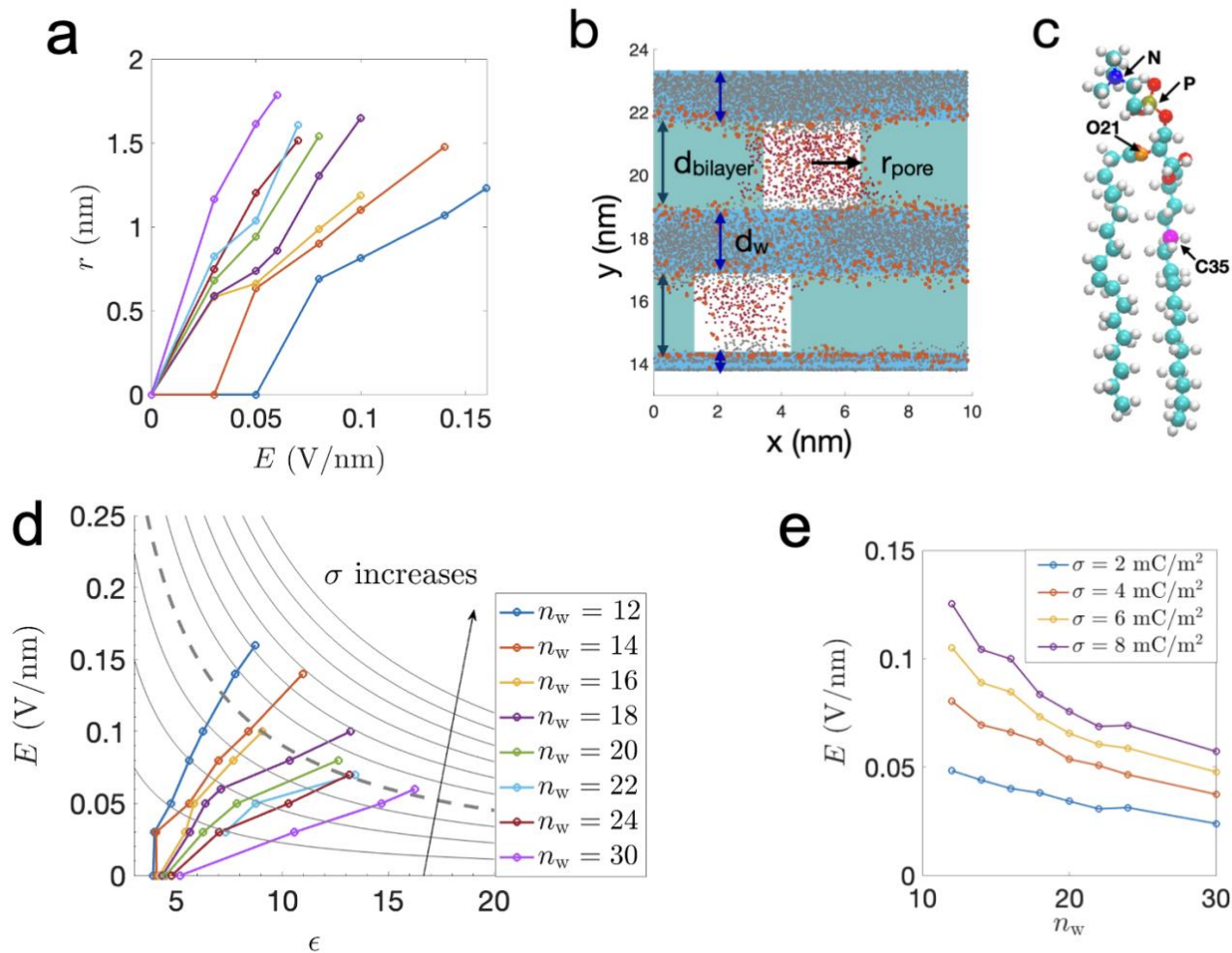

Figure 2. a) Mean pore radius at equilibrium for varying electric field and hydration levels (for colour-coding of different hydration levels see box in fig. 2d). b) The structural parameters determined by the analysis. $d_{\text{bilayer}}$ and $d_w$ are respectively the summed thickness of the lipid phase and the water phase. c) Structure of an all-atom POPC: C (cyan), H (white), O (red), N (blue), P (gold). Atoms relevant to the analysis are specifically labeled. d) Directly applied electric fields versus dielectric constant evaluated based on the structural parameters of each system. Grey lines are iso-surface charge density curves from $\sigma$ = 2 mC/m² to 20 mC/m² at 2 mC/m² intervals evaluated using Gauss's law $E = \sigma/\epsilon\epsilon_0$, where $\epsilon_0$ is vacuum permittivity. The bold dashed curve corresponds to the SFB experiments. e) The equilibrium electric field across bilayers confined between two charged surfaces with opposite and equal surface charge.

Given the same external electric field, larger pores are found with more hydrated membranes, while a threshold electric field is found for $n_w$=12 and 14 at ca. 0.03-0.05 V/nm, as the 0.5 V/nm-pre-initiated pores close at this field. Notably, electropores are maintained by similar electric fields with higher hydration levels (Figure S3).

As the membrane becomes more porated (larger pore diameters at higher $E$, fig. 2a), the mean dielectric constant $\epsilon$ of the entire system increases, given the large differences between the dielectric constants of lipid and of water as given above. By identifying the boundaries between the water phase and lipid phase in the simulations (Figure 2b), $\epsilon$ is evaluated (see Analytical Methods). This provides a relation between $E$ and $\epsilon$ for different hydration levels (Figure 2d). For self-consistency, the $E(\epsilon)$ relation must also obey Gauss's law $E = \sigma/\epsilon\epsilon_0$, where $\sigma$ is the surface charge density on the surfaces confining the bilayers and $\epsilon_0$ is the vacuum permittivity. This additional condition enables us to evaluate the field to be used in the simulations with reference to the experiments. In the SFB experiments, the mica surface in the water bath is negatively charged due to the dissociation of surface potassium ions, which results in a surface charge density of ca. -8 mC/m$^2$ [49], while the gold surface is held at a constant potential in the order of 0.1-0.3 V relative to the reference electrode in the bath. As the bilayer-bearing surfaces approach, the charge density on gold reverts to be opposite and approximately equal to that of mica at a separation $D$ of two-bilayer thickness of ca. 10 nm, due to charge inversion described by Poisson-Boltzmann theory, as previously discussed [49, 58]. We therefore model the simplistic situation where the surface charge on gold is 8 mC/m$^2$ with no ions in the mica-gold gap, as it is entropically favorable for them to escape to the bulk (Figure S 4). We then plot the iso-charge curves of Gauss's law for varying $\sigma$ in Figure 2d, and the solutions of external electric field for any given surface charge is found at the intercepts of the two sets of curves: the $E(\epsilon)$ relations derived from the simulations (figs. 2a, 2b) and the Gauss's law $E = \sigma/\epsilon\epsilon_0$. The results for the varying hydration levels of the lipids, $E(n_{\text{water}})$ in Figure 2e, show that the electric field for a fixed surface charge decreases as hydration level increases and converges to values of O(0.01) V/nm at $n_w$=30. This magnitude is consistent with a previous experimental study of a water-only system with the identical SFB set up, where the electric field evaluated at a separation of 10 nm is c.a. 0.012 V/nm [49]. From the plot, the electric fields for $\sigma$=8 mC/m$^2$ (relevant for the SFB experiments) for the two relevant hydration levels $n_w$=12 and 20 are 0.13±0.01 V/nm and 0.08 ±0.01V/nm respectively, which guide our choice of a field $E$ = O(0.1 V/m) in the simulations. The corresponding porated structures are then repeated in four independent simulations.

## 2.2 Lipid merging induced by an electric field of O(0.1) V/nm

The configuration of our simulations and representative electroporated structures arising from application of the electric field across the two-bilayer stack (either two-bilayer-only or confined by gold and mica surfaces) are shown in fig. 3. For the $n_w$ = 12 system, the repeated porated structures are all structurally similar (Figure S5). However, for $n_w$ = 20, a unique structural feature, not previously reported but of considerable interest as noted below, is observed in one of the four repeats: A lipid bridge self-assembles, penetrating through the midplane water phase and connecting the two bilayers (Figure S5). As this structure seems to be affected by the finite size in x, y-directions, we also carried out matching simulations with coarse grained models, where the planar area is quadrupled. Among the five repeats, two structures with lipid bridge are observed, confirming that it is a less-frequent but regularly occurring effect (Figure 3a).

Smirnova et al. (2010) has demonstrated that the rate-limiting step for lipid merging is the hydrophobic contact between the two bilayers in the form of splayed lipid bonds, based on which we speculated that bridge formation by electric field can be attributed to two effects: 1) facilitated dehydration through electroporation and consequently reduced hydration repulsion at local contact points, and 2) disturbed lipid packing order and promoted membrane surface fluctuation, which increases the rate of an inter-membrane hydrophobic contact. The fact that the bridge formation is observed at the higher hydration level $n_w$=20 should not be surprising, as existing studies demonstrated that water defects on supported membranes contributes to flip-flop rates higher by orders of magnitude comparing with liposomes of defect-less lipid packing [59-61]. In addition, a higher hydration level corresponds to higher lipid diffusion rate [62]. Here we see the duality of hydration effects—while hydration repulsion keeps membranes well separated, it locally increases the rate of lipid-lipid interaction with elevated lipid thermal motion. The later effect is especially pronounced when an electric field introduces topological disorder.

We associate the formation of lipid bridge with significant, topologically-forced effects on the frictional dissipation when the two confining bilayer-bearing surfaces slide past each other [9], and for this reason it is chosen for the next steps of analysis with solid boundaries. For these, two types of simulations were performed: the charge-imbalance simulations which retain the electric field with charged surfaces, and simulations to test reversibility with uncharged surfaces (Figure 3 b-c).

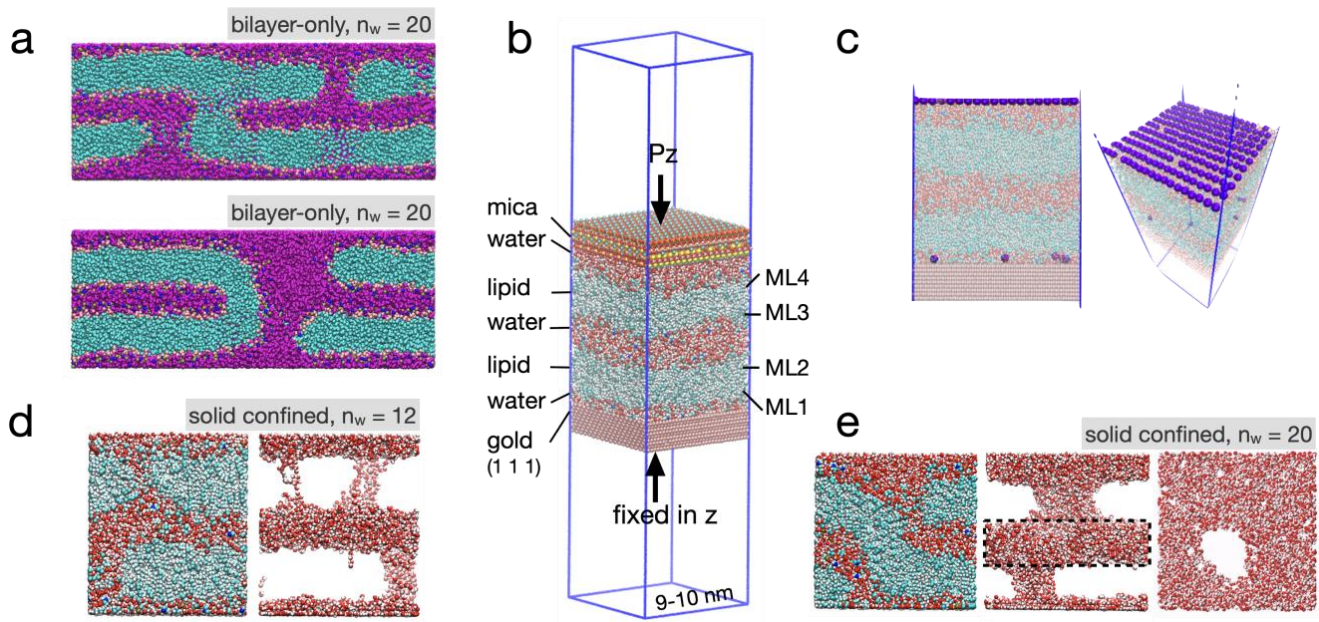

Figure 3. a) Side cross-section views of selected coarse-grained equilibrium simulations for $n = 20$ with $E \approx$ 0.1 V/nm, demonstrating sporadically occurring lipid bridge. Purple: coarse-grained water molecule. Turquoise: coarse-grained bilayer. b) Snapshot of the planar simulation system of periodical replication in x and y. For the equilibrium simulations, the z-coordinates of gold atoms are restrained, and pressure is imposed at the center of mass of the mica slab. Mica: K (cyan), Si (yellow), Al (pink), O (red). Water: O (red), H (white). Lipid: C (cyan), H (white). Gold: pink. c) Charge-imbalance imposed by transferring surface potassium atoms (violet) from mica (not shown) to gold (lower) surface. The left-hand side shows the potassium ions being transferred to the bottom. The right-hand side shows the missing potassium ions after the procedure. d) Side cross-section views of equilibrated electroporated membrane structure of $n_w$=12, where solids are not shown. Right: lipids removed to reveal water pores. e) Left and middle: side cross-section views of electroporated membrane structure of $n_w$=20. Middle: Lipids removed to reveal water pores. Right: top-view of the midplane interbilayer water phase indicated by the box of the middle panel, revealing a nanometric cylindrical lipid bridge.

## 2.3 Structures of porated membranes under confinement and reversibility

Headgroup reorientation.

The all-atom pore structures for both hydration levels are shown in Figure 3 d-e. The most significant effect of the solid boundaries is that membranes conform to the flat solid surfaces (Figure S 5). This affects the well-known toroidal shaped electropores formed by the hydrophilic headgroups of the lipids [11, 15, 63]. Distributions of the angles of P-N vectors (of the phosphocholine groups) relative to the +z direction $\theta$ of each monolayer (ML) are plotted in Figure 4 a. With both hydration levels, the monolayer in immediate vicinity to the gold surface (ML1) has headgroups aligned parallel to the gold surface with a narrow distribution around 89°, due to the strong Van der Waals interaction with the gold, a highly dense solid [64]. Mica-side headgroup orientations (ML4) are less restricted by the mica surface, and at the higher hydration level of $n_w$=20, the distribution is almost identical to the midplane monolayer ML2. With both hydration levels, the mean headgroup orientations of the midplane monolayers, ML2 and ML3, decrease from 69° and 110° by 3° and 10° respectively as the anodic and cathodic zwitterionic P-N headgroups respond to the electric field unsymmetrically. This is consistent with previous analysis that even under unconfined conditions, the shifts in headgroup orientation distribution is only a few degrees under a field of 0.1 V/nm due to tight lipid packing [10]. This shift is expected to have negligible effects on the bilayer-bilayer interaction, as the distribution of $\theta$ spans a wide range with a standard deviation of 29°.

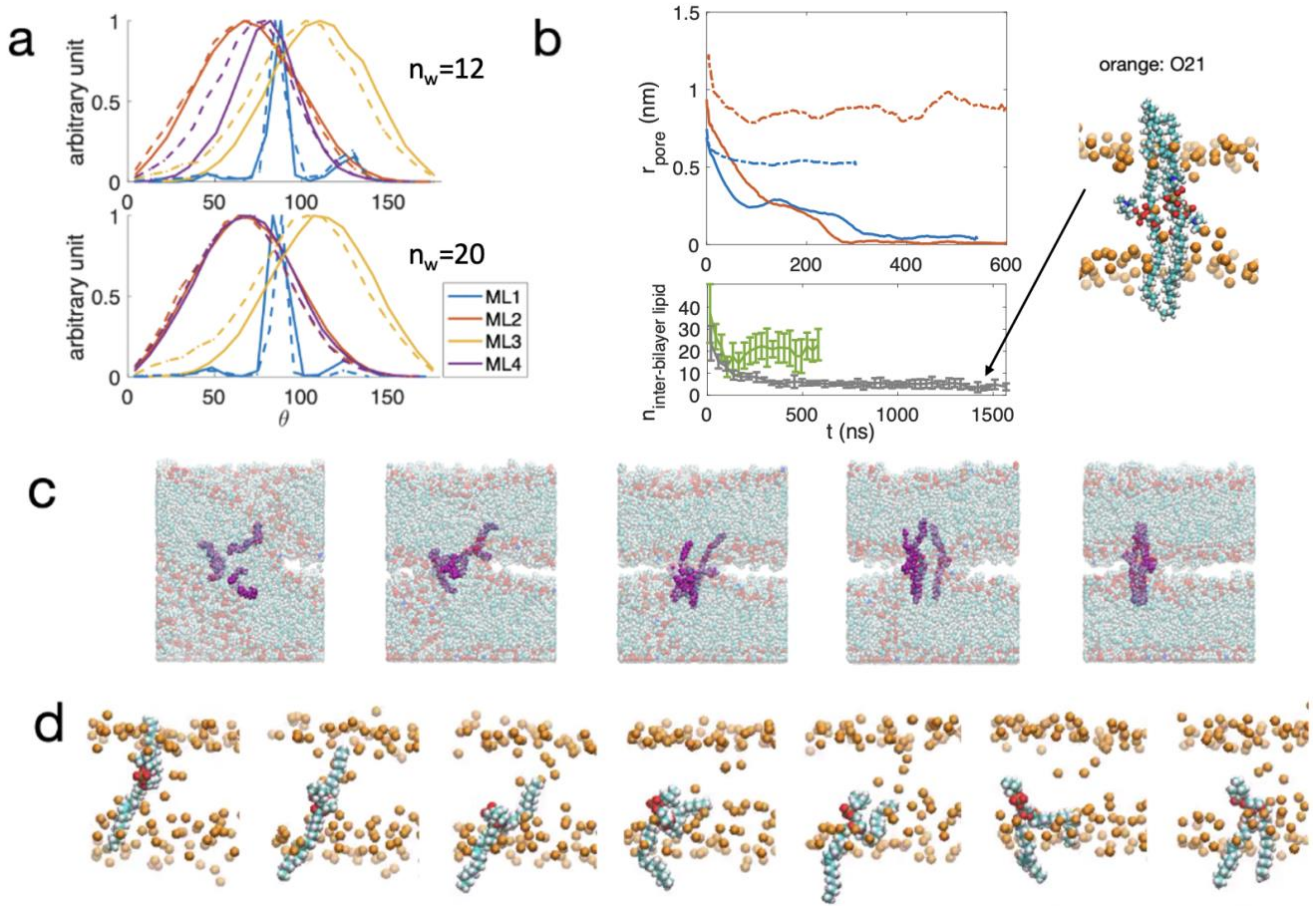

Figure 4. Structural analysis of confined bilayer systems. a) Distributions of P-N angles for the four monolayers (ML). Top: $n_w$=12. Bottom: $n_w$=20. Solid lines: $E$=0. Broken lines: $E≈0.1$ V/nm. b) $r_{pore}$ (top) and $n_{interbilayer\ lipid}$ (bottom) as a function of equilibration time after the addition of the solid slabs to the membrane-only system. Top: Blue: $n_w$=12. Orange: $n_w$=20. Solid lines: $E$=0 V/nm. Broken lines: $E≈0.1$ V/nm. Bottom: Green: $n_w$=20, $E≈0.1$ V/nm. Grey: $n_w$=20, $E$=0 V/nm. For the case of $n_w$=20, $E$=0, $n_{interbilayer\ lipid}$ does not reduce to zero due to three splayed lipids molecules (snapshot). c) Retraction of lipid bridge under $E$=0 ($n_w$=20). The three lipids which remained splayed across the membranes at $t$=1.6 μs are marked as purple. d) A splayed lipid transformed back to the lamellar state with a duration of 400 ns.

Pore and bridge structures and reversibility.

For $n_w$=12, the mean water pore radius decreases from 0.7 nm (for the unconfined membranes) to 0.5 nm as the undulations caused by the toroidal pores are suppressed upon confinement. Likewise, the $n_w$=20 system pores reduced from a mean radius of 1.2 nm to a decreased value of 0.8 nm on confinement. Similarly, the lipid bridge of the $n_w$=20 system stabilizes at a decreased diameter of 3 nm in response to the solid surfaces (Figure 4b). Finally, reversibility is demonstrated by the closure of all pores in both $n_w$=12 and $n_w$=20 systems within 500 ns when the electric field reverts to $E$=0 (SI video 1). For $n_w$=20, however, the lipid bridge reduces to sub-nanometer diameter but does not close within 1.6 μs (Figure 4 b, SI video 2). When the electric field reverts to zero, the lipids self-rearrange to integrate back into the bilayers. Within the times of our simulations (up to 1.6 μs), however, some lipids remained splayed across the two bilayers, with the hydrophilic headgroups in the inter-membrane water phase and the two tails residing in the acyl tail region (Figure S 6). This arrangement is metastable especially when the water phase is thin, where the contact between the acyl tails of the splayed lipids and the water molecules is minimized (Figure 4 c). The splayed lipids can transform back into the lamellar phase, as one such case is seen at $t$=960 ns (Figure 4 d, SI video 3). When tested with the coarse gained simulations, both the bridge and the pores disappear within 100 ns. In addition, Smirnova et al. (2010) has demonstrated with coarse grained simulations that even at extremely low hydration of $n_w$=5, splayed POPC lipids can sporadically transform back to the lamellar state [36]. It is therefore reasonable to assume that the lipid bridge will entirely disappear in the all-atom simulations if equilibration is extended by additional microseconds.

## 3. Summary and Conclusions

The main new findings of this study, where atomistic MD simulations were carried on bilayer stacks confined between two solid surfaces, is to demonstrate reversible, hydration-dependent membrane restructuring, and to elucidate the nature of this restructuring, when a transverse electric field of order of ca. 0.1 V/nm was applied across the intersurface gap. This is close to the field applied in the SFB experiments whose results [9], showing massive modulation by the field of the sliding friction between surfaces across lipid bilayers, were a motivation for this study. The hydration levels were estimated realistically as a range defined by the minimum hydration level of the hydration shell and the elevated hydration level due to water defects. Electroporation is observed at all hydration levels. At higher hydration levels, a lipid bridge can form, resembling the so-called stalk structure, an intermediate metastable state of hemifusion [39]. The formation of such bridging structures under an electric field has not as far as we know been previously reported.

In sum, motivated by recent SFB experiments on confined lipid bilayers under electric fields [9] we investigated two types of systems — infinite (unconfined) lipid membrane stacks implemented through periodic boundary conditions and a membrane stack confined between solid slabs. This is the first time that the effects of electric fields on such stacked or confined lipid bilayers have been examined by atomistic simulations. We took care to design the protocol of solid confinement to ensure that electric-field-induced lateral expansion of the membranes would not be restricted by the finite size of the solid phase. The results demonstrate that the field-induced structural/topological changes at the membrane-membrane interface are similar both for stacked but unconfined membranes, as well as in the presence of confining surfaces, even for strong solid-membrane interaction in the latter case, indicating the generality of the effects. We also took care to work at electric fields and confining slabs similar to those in the surface forces experiments, and our results provide insight, explored in more detail elsewhere[42], into the strong effects of electric fields on the interactions, particularly shear and frictional interactions, between bilayer-bearing surfaces.

**4. Results and discussion**

4.1     Equilibrium simulations of bilayer-only systems

4.1.1   All-atom simulations

The simulation parameters were adopted from existing lipid electroporation MD simulation studies [20, 21, 23, 56, 65]. The initial simulation system comprises two 1-palmitoyl-2-oleoylphosphatidylcholine (POPC) bilayers with 128 lipids per monolayer leaflet and with varying hydration levels at 12, 14, 16, 18, 20, 22, 24 and 30 water molecules per lipid, built by the CHARMM-GUI generator [66-69]. We used the three-site TIP3P water model [70], and the OH-bonds are constrained with SETTLE constraints [71]. For the lipids we used the CHARMM36 all-atom parameters [72]. Bond lengths were constrained using the LINCS algorithm [73]. Lennard-Jones interactions were cut off at 1.0 nm. The electrostatic interactions were treated with the particle-mesh-Ewald (PME) method [74, 75] with a grid spacing of 0.12 nm, a fourth-order spline, and a real-space cut-off distance of 1.1 nm. All simulations were executed with the GROMACS package. The simulation time step is 2 fs. The system was first equilibrated in a NPT ensemble, where periodic boundary condition is applied in all directions. A velocity-rescaling thermostat was applied with T=298 K and a time constant of 0.1 ps. Berendsen barostat was applied with anisotropic pressure-coupling, where $P_{x,y}$ =1 bar and $P_z$ = 10 bars, similar to the experimental condition. The time constant is 2.5 ps and the compressibility is $4.5 \times 10^5$ bar$^{-1}$. The system is equilibrated for 100 ns at zero electric field. To generate a stable electropore, we followed the protocols in existing literature where a higher electroporating electric field of the magnitude 0.5 V/nm normal to the membrane was applied using the direct electric field method, where a force, evaluated as the product of the partial charge of an atom and the field strength is imposed onto each atom [22, 57]. Once an electropore is induced, stabilizing electric fields of various magnitudes between 0.03 V/nm to 1.60 V/nm is applied and equilibrated under the same NPT conditions for 400 to 1000 ns until the area per lipid and pore radius plateaus [10, 12, 22, 57]. This procedure is to accommodate the limited computational time of $O(1)$ μs at maximum, whereas in the experiments, pores take longer time to form under low electric fields.

### 4.1.2 Coarse-grained MARTINI simulations

MARTINI simulations of large membrane size were carried out for $n_w$=20 to examine equilibrated electroporation structures at 0.08 V/nm. The simulation systems comprise two 1-palmitoyl-2-oleoylphosphatidylcholine (POPC) bilayers with 512 lipids per monolayer leaflet built by the CHARMM-GUI generator with the martini22p forcefield, with Martini 2.0 lipids and the polarizable water model with each unit representing four water molecules [29, 66, 76-78]. The neighbor list is updated every ten steps using the Verlet neighbor search algorithm. Bond lengths were constrained using the LINCS algorithm [73]. Lennard-Jones potential was shifted to zero between 0.9 and 1.2 nm and electrostatic interactions between 0.0 and 1.2 nm. The long-range electrostatic interactions were calculated with the particle-mesh-Ewald (PME) method. The relative electrostatic screening is 2.5 as recommended for polar water model. A velocity-rescaling thermostat was applied with T=298 K and a time constant of 1 ps. Parrinello-Rahman barostat was applied in NPT simulations with time constant of 12 ps and compressibility of $3 \times 10^{-4}$ bar$^{-1}$ [74, 75]. The simulation time step is 20 fs. Periodic boundary condition is applied in all directions. To achieve consistency with the all-atom simulation for direct comparison, the lipid-water systems were first treated with a higher electric field of 0.5 V/nm under NPT conditions until the area per lipid matches that of the equilibrated porated all-atom structure at 0.08 V/nm. Then the simulation box is fixed at NVT condition and because poration has not fully occurred at this stage, an electric field of 0.5 V/nm was resumed until the membranes are porated. Then the electric field is reduced to 0.08 V/nm and equilibrated at NVT condition for 100 ns.

## 4.2 All-atom equilibrium simulations of membranes with planar solid confinement

### 4.2.1 Charge-imbalance simulations of E ≈ 0.1 V/nm systems

The membrane-only structures of $n_w$=12 and $n_w$=20 equilibrated at an electric field of 0.14 V/nm and 0.08 V/nm respectively were equilibrated again after the addition of the solid phase with the following procedures. Solid slabs of mica and gold (111), both generated using the CHARMM-GUI generator based on INTERFACE FF [79], were placed respectively above and below the equilibrated membranes. The thickness of both solid slabs is 2 nm, which was tested to be sufficiently rigid under the applied pressure. In theory, the area of the solid slabs should accurately match that of the equilibrated membrane structure. This is difficult for mica due to its finite size crystal lattice, respectively 5 Å and 9 Å in x-y directions. An areal difference of 3% was unavoidable. Finally, vacuum space was added to the top and bottom of the structure to reduce interaction between the solid phases across the periodic boundary in z-direction. In SFB experiments, mica and gold surfaces are charged with c.a. $\mp 8$ mC/m$^2$ at a separation of ~10 nm (see main text) [49]. For our double-bilayer systems of 80-100 nm$^2$ in cross-section area (i.e. the area after the application of electric field in absence of solids), it is equivalent to a surface charge number of 4.5-5.6. Practically, we translate five potassium atoms from randomly scattered positions at the inner mica surface to the inner surface of gold to match the experimental condition (Figure 2 c). During the equilibrium simulations, positions of the gold surface and the five surface potassium charges are restricted in z to maintain the charge imbalance. This step is designed such that the increase in cross-section area due to poration is included when the non-stretchable solid phases are incorporated, while the charge-imbalance treatment imposed is consistent with the electric field previously imposed to the membrane-only system.

### 4.2.2 Testing pore closure with E =0 V/nm systems

To test the reversibility of electroporation with solid confinement, the membrane-only structures of $n_w$=12 and $n_w$=20 equilibrated with an electric field of 0.14 V/nm and 0.08 V/nm respectively were also equilibrated with solid confinement without imposing charge imbalance. The solid slabs of gold and mica were prepared to match the cross-section area of $n_w$=12 and $n_w$=20 membrane-only systems equilibrated under zero electric-field.

All solid-membrane systems were equilibrated under matching conditions with the membrane-only system. The box height is fixed by setting the isothermal compressibility in z to be 0 bar$^{-1}$, which prevents the vacuum space from collapsing. A normal constant compression force of 53 kJ

mol$^{-1}$nm$^{-1}$ (≈ 88 pN) is applied to the center of mass of the mica slab, resulting in a pressure of 10±1 atm for equilibrated surface areas varied from 79 nm$^2$ to 98 nm$^2$, (corresponding to 12 water/lipid under zero electric field and 20 water/lipid under ~0.1 V/nm electric field respectively). The structure (pore size, interbilayer lipids, cross-section area) were monitored for a duration of 300-1600 ns until convergence.

**Acknowledgements**

We thank the McCutchen Foundation, the Israel Science Foundation-National Natural Science Foundation of China joint program (ISF-NSFC Grant 3618/21), the WIS - UChicago - TTIC - AI+Science Collaborative Research Program, and the Israel Science Foundation (Grant ISF 1229/20) for their support of this work. This project has received funding from the European Research Council under the European Union's Horizon 2020 research and innovation programme (Grant Agreement 743016). This work was made possible in part by the historic generosity of the Harold Perlman family.

# Supplementary materials

# Analytical methods

**Evaluating pore radius and dielectric constant.** Coordinates of the O21 atoms (Figure S 7 a) were exported from the GROMACS trajectories, and the distribution of the atoms along the membrane normal is acquired using MATLAB scripts. Gaussian distribution was fitted to extract the z-coordinates of the four peaks $z_i$ *and* standard deviation $\delta_i$ (*i*= ML1-4) which identifies the boundaries between the lipid phase and the water phase (Figure S 7Figure S *7* b). The number of water molecules in pore $n_{pore}$ is evaluated by counting the number of water molecules which satisfy $z_{ML1} + \delta_{ML1} < z_w < z_{ML2} - \delta_{ML2}$, or $z_{ML3} + \delta_{ML3} < z_w < z_{ML4} - \delta_{ML4}$. The number of water molecules in the continuous water phase are therefore $n_{\text{water slab}} = n_{tot} - n_{pore}$. The total thickness of the continuous water phase is evaluated as $d_w = n_{\text{water slab}} M_w/(N_A \rho_w A)$, and the total thickness of the lipid phase is evaluated as $d_{\text{bilayer}} = d_{\text{tot}} - d_w$, where the cross-section area $A = (x_{\max} - x_{\min})(y_{\max} - y_{\min})$, and $d_{\text{tot}} = z_{\max} - z_{\min}$ are evaluated with the coordinates of the outmost water molecules at system boundaries. The mean pore radius is then evaluated as $r_{\text{pore}} = \sqrt{\frac{V_{\text{pore}}}{\pi\, d_{\text{bilayer}}}}$, where $V_{\text{pore}} = n_{\text{pore}} M_w/(\rho_w N_A)$. The dielectric constant is then estimated as $\epsilon = \frac{\epsilon_{\text{bilayer}}\, \epsilon_w (d_{\text{bilayer}} + d_w)}{\epsilon_{\text{bilayer}}\, d_w + \epsilon_w\, d_{\text{bilayer}}}$, and $\epsilon_{\text{bilayer}} = \epsilon_{\text{lipid}} \left(1 - \frac{A_{\text{pore}}}{A}\right) + \epsilon_w \frac{A_{\text{pore}}}{A}$, respectively corresponds to permittivity of materials arranged in series and in parallel. $\epsilon_w = 80$ and $\epsilon_{\text{bilayer}} = 3$ are used [1]. In the special case of porated membranes with $n_w=20$ with interbilayer lipids, the 10% decrease in the permittivity of the interbilayer water phase is neglected.

**Evaluating number of interbilayer lipids.** The number of lipids found between the bilayers intruding through the mid-plane water phase is evaluated from the O21 distribution. The number of the O21 atoms which satisfy $z_{ML2}+2\sigma_{ML2} < z_{O21} < z_{ML3}-2\sigma_{ML3}$, where $\sigma$ is the standard deviation of the Gaussian distribution fitted, is evaluated from snapshots (Figure S 8. Distribution of O21 atoms from the two inner monolayers of nw=20 with E ≈ 0.1 V/nm. The grey boundaries are evaluated as zML2+2σ and zML3-2σ. The total number of O21 atoms between the two boundaries are evaluated, representing the number of inter-bilayer lipids.

).

**P-N angle analysis.** Coordinates of the C35 atoms were exported from the GROMACS trajectories, and four Gaussian distributions were fitted. The lipid molecule indices were assigned to each of the four monolayers based the position of its C35 atom (Figure S 7 a, c). The coordinates of P and N atoms were then extracted for each monolayer and the angle of P-N vector relative to +z is evaluated as $\theta = \cos^{-1}(d_{PN}/d_z)$, where $d_{PN} = |\vec{P} - \vec{N}|$ and $dz = P_z - N_z$

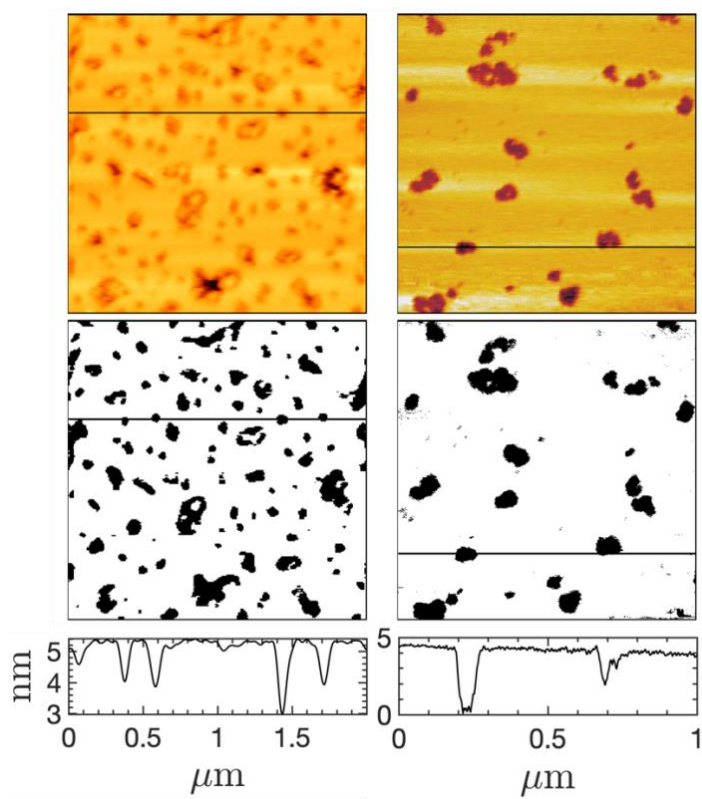

Figure S 1. Estimating area fraction of water defects from AFM scans of POPC bilayers on mica surfaces using image segmentation. The area fraction of water defects are respectively 15.8% and 9.8% for the scans on the left and right. Bottom: depth profiles corresponding to the cross-section indicated in the scan image.

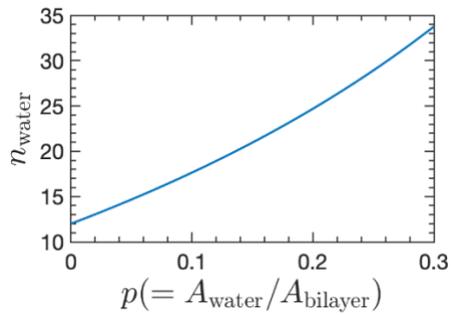

Figure S 2. Number of water molecules per POPC molecule $n_{water}$ evaluated for given area fraction of water defects $p$, assuming $n_{water}(p=0) = 12$ for perfect POPC bilayer confined and pressurized by solid boundaries. $n_{water} = n_{water}(p=0) + \frac{p}{1-p} apl\, H_0 \rho_w/(M_w NA)$, where $apl = 0.62\ nm^2$ is the area per lipid evaluated from the equilibrium state of the $n_{water} = 12$ double-bilayer simulation under zero electric field, $H_0 = 5$ nm is the thickness of a monolayer, $\rho_w$ is the water density, $M_w$ is the molecular weight of water, $NA$ is the Avogadro's number

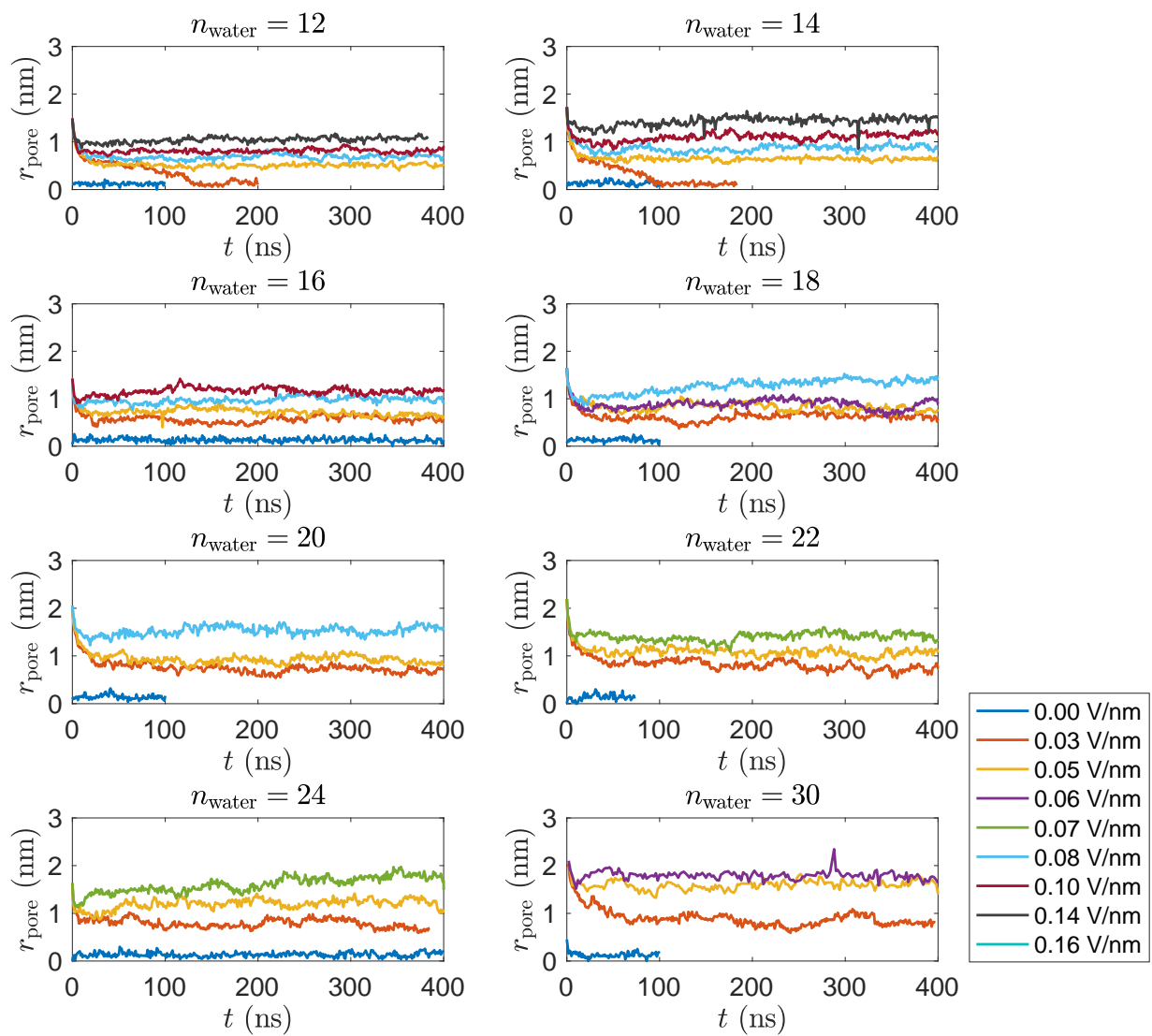

Figure S 3. Water pore radius evaluated during the equilibration process at varying electric fields.

The title of axis and each panel are little bit blurry.

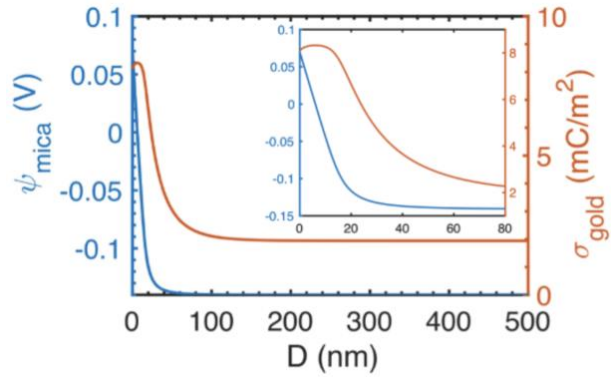

Figure S 4. Surface potential of mica $\psi_{mica}$ and surface charge of gold $\sigma_{gold}$ solved for the SFB experimental configuration [2].

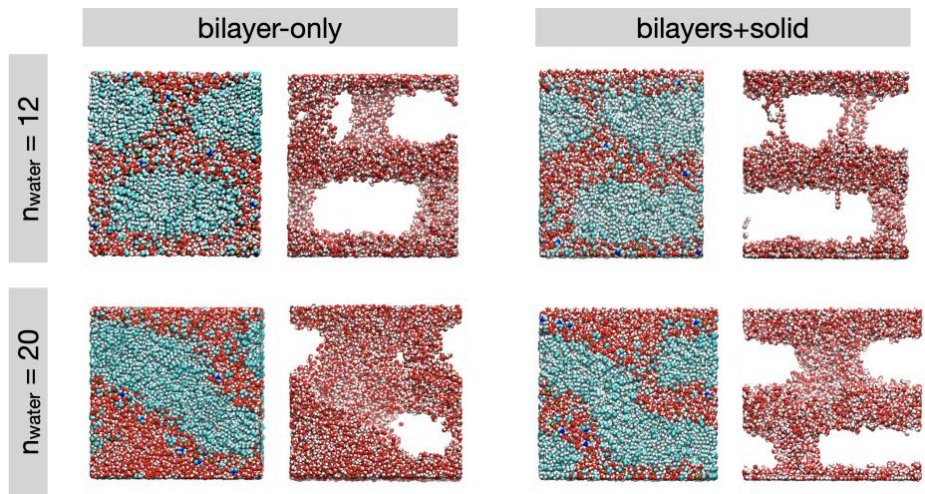

Figure S 5. Equilibrated electroporation structures at E≈0.1 V/nm. The solid slabs are ignored for the clear illustration of lipid-structures in the right column.

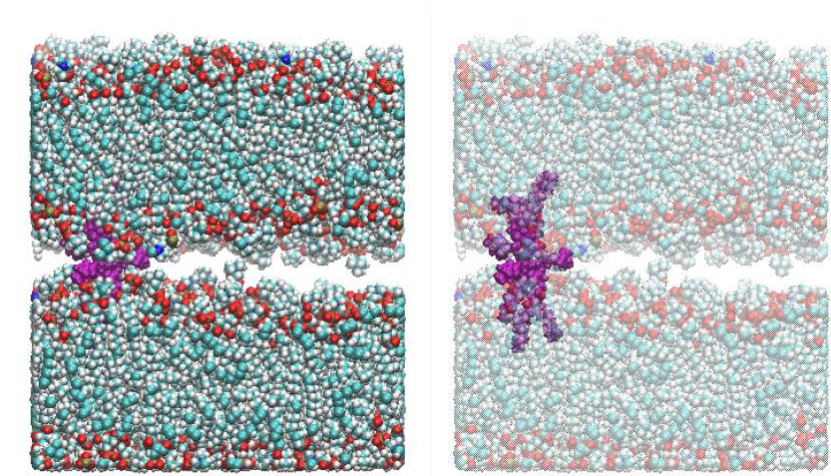

Figure S 6. Reduction of lipid bridge after electric field is reverted to $E=0$ ($n_w$=20, t=1.6 μs). The three purple lipids splayed across the inter-membrane water phase are metastable, with estimated life time of O(1) μs before being entirely integrated into either bilayer.

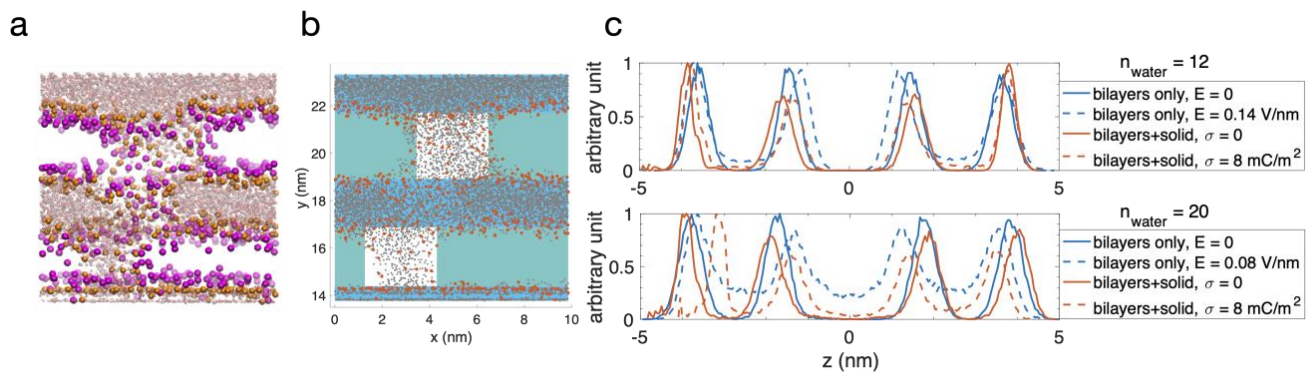

Figure S 7. Analysis with C35 and O21. a) Snapshot of porated membrane. Red and white semi-transparent molecules: water. Orange: C35. Pink: O21. b) Pore region (white), water phase (light blue) and lipid phase (turquoise) identified by analysis. Orange: O21. c) Distribution of C35 for different systems.

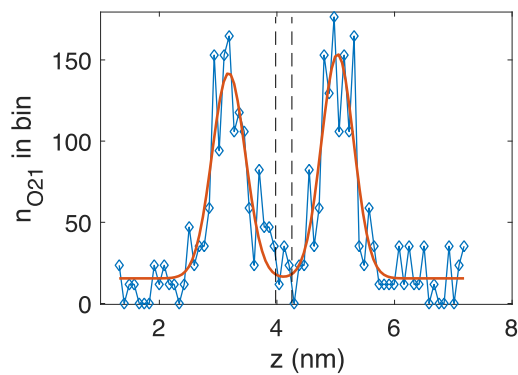

Figure S 8. Distribution of O21 atoms from the two inner monolayers of nw=20 with E ≈ 0.1 V/nm. The grey boundaries are evaluated as zML2+2σ and zML3-2σ. The total number of O21 atoms between the two boundaries are evaluated, representing the number of inter-bilayer lipids.